\documentclass[twocolumn]{aa} %
\usepackage{natbib,graphicx}
\newcommand{\Mjup}{M$_{\rm Jup}$}

\newcommand{\au}{{\sc au}}
\newcommand{\betapic}{$\beta$~Pic}

\begin{document}
   \title{Sparse aperture masking at the VLT}

   \subtitle{I. Faint companion detection limits for the two debris
     disk stars HD\,92945 and HD\,141569}

   \author{S. Lacour\inst{1} \and
          P. Tuthill\inst{2} \and
          P. Amico\inst{3} \and
          M. Ireland\inst{2} \and
          D. Ehrenreich\inst{4} \and
          N. Huelamo\inst{5} \and
          A.-M. Lagrange\inst{4}}

   \institute{LESIA, CNRS/UMR-8109, Observatoire de Paris, UPMC, Universit\'e Paris Diderot, 5 place Jules Janssen, 92195 Meudon, France
\\
              \email{sylvestre.lacour@obspm.fr}
         \and
         Sydney Institute for Astronomy, School of Physics, The University of Sydney, NSW 2006, Australia
             \and
             ESO, Karl-Schwarzschild-Strasse 2, 85748 Garching, Germany
         \and
UJF-Grenoble 1 /  CNRS-INSU, Institut de plan\'etologie et d'Astrophysique de Grenoble (IPAG) UMR 5274, Grenoble, F-38041, France
         \and
         Centro de Astrobiolog\'{\i}a (INTA-CSIC);  LAEFF, P.O. Box 78, E-28691 Villanueva de la Ca\~nada, Spain\\
             }

   \date{}

 
  \abstract
{}  
   {Observational data on companion statistics around young stellar systems is needed to flesh out the formation pathways for extrasolar planets and brown dwarfs. 
   Aperture masking is a new technique that is able to address an important part of this 
   discovery space.}
{We observed the two debris disk systems HD\,92945
   and HD\,141569 with sparse aperture masking (SAM), a new mode offered on the 
   NaCo instrument at the VLT.
A search for faint companions was performed using a
detection strategy based on the analysis of closure phases recovered
from interferograms recorded on the Conica camera.}
{Our results demonstrate that SAM is a
very competitive mode in the field of companion detection.  We
obtained $5\sigma$ high-contrast detection limits at $\lambda/D$ of
$2.5\times10^{-3}$ ($\Delta L' = 6.5$) for HD\,92945 and
$4.6\times10^{-3}$ ($\Delta L' = 5.8$) for HD\,141569. According to
brown dwarf evolutionary models, our data impose an upper mass
boundary for any companion for the two stars to, respectively, 18 and
22 Jupiter masses at minimum separations of 1.5 and 7~\au. The
detection limits is mostly independent of angular separation, until
reaching the diffraction limit of the telescope.}
   {We have placed upper limits on the existence of companions to our
     target systems that fall close to the planetary mass regime.
     This demonstrates the potential for SAM mode to contribute to
     studies of faint companions. We furthermore show that the final
     dynamic range obtained is directly proportional to the error on
     the closure phase measurement.  At the present performance levels
     of 0.28~degree closure phase error, SAM is among the most
     competitive techniques for recovering companions at scales of
     one to several times the diffraction limit of the telescope.
     Further improvements to the detection threshold can be expected
     with more accurate phase calibration.}


   \maketitle
%

\section{Introduction}

   \begin{figure*}
   \centering
   \includegraphics[scale=.45]{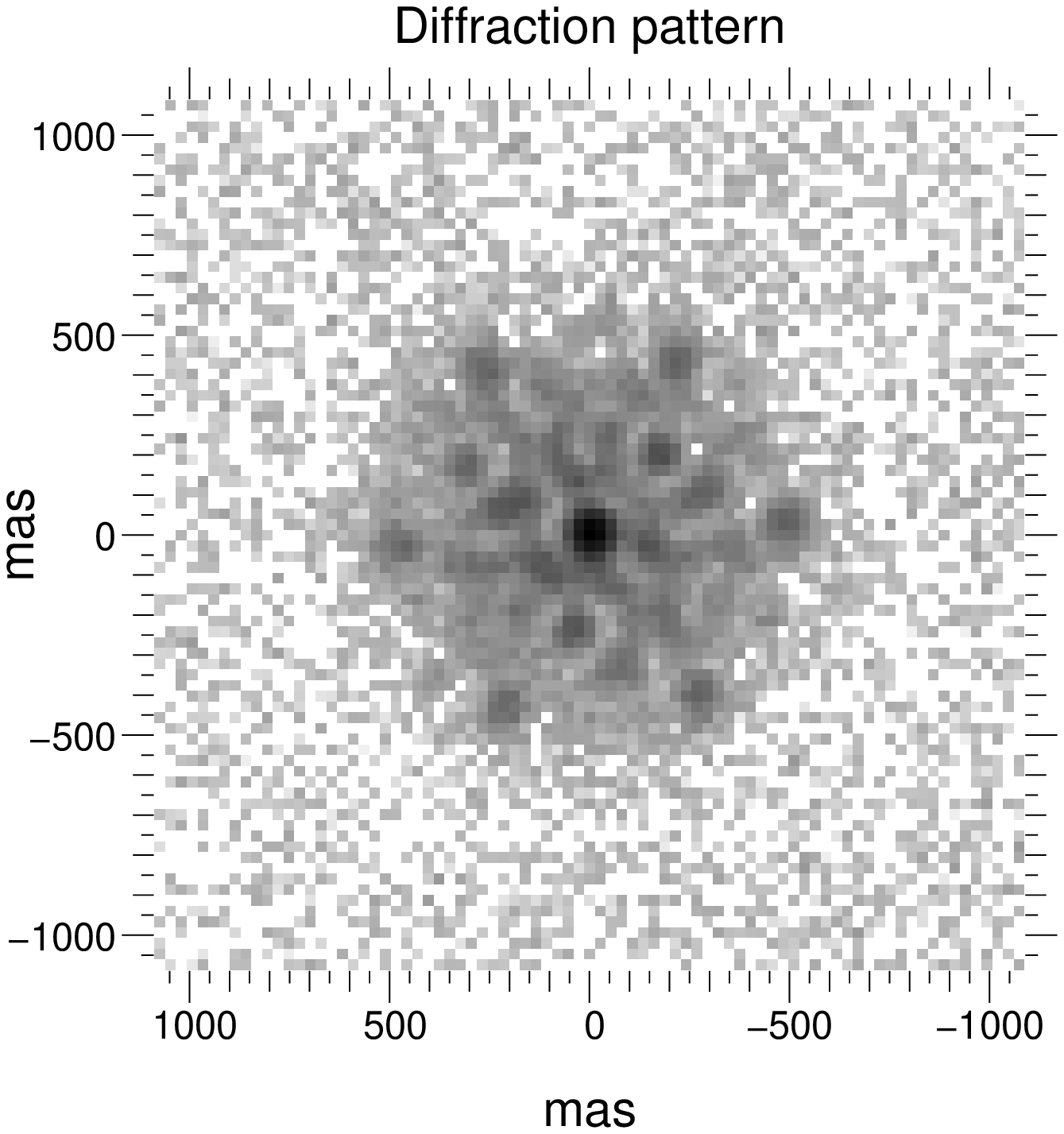}\hspace{1cm}
   \includegraphics[scale=.45]{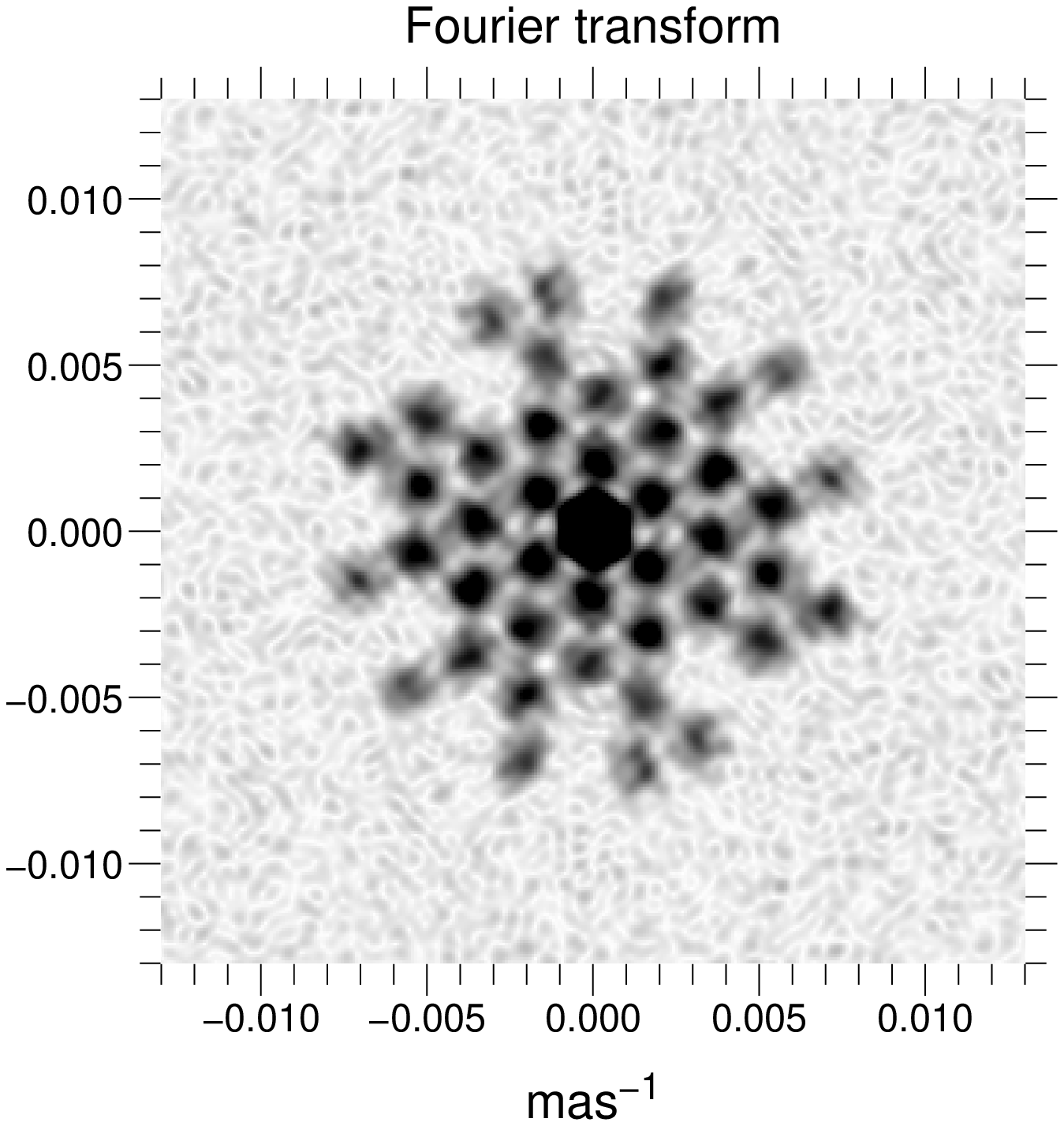}
   \caption{Left panel: Diffraction pattern observed on star
     HD\,135549. This image is a single frame of 0.4\,ms integration
     time, using the 7-hole pupil mask. Background noise is $\approx$12 ADU,
     and pic brightness 800 ADU. Right panel: Fourier transform of
     the diffraction pattern. Each peak corresponds to a baseline
     pair.}
              \label{Image}%
    \end{figure*}

Since the first imaging of a debris disk around $\beta$~Pictoris
\citep{1984Sci...226.1421S}, and the remarkable successes of the
radial velocity technique for exoplanet discovery heralded nearly a
decade later by a companion to solar-like 51~Pegasi
\citep{1995Natur.378..355M}, the understanding of planetary systems
formation and evolution has become one of astronomy's greatest
challenges.  Today, at least 22 debris disks have been imaged at
optical, infrared, or submillimeter wavelengths.  Such disks are
thought to be the cradle where planetary systems have recently
formed. The observed dust component is not primordial but instead
produced by collisions among larger rocky bodies. This makes debris disks
 obvious places to search for planets, and many show possible indirect signs
of their presence, such as dust structures or disk asymmetries.

The recent detection and confirmation of a $\sim9$~\Mjup\ planet
within the disk of \betapic\ (semi-major axis between 8 and 15~\au)
using direct deep imaging validates the link between structures in
debris disks and the presence of planets
\citep{2009A&A...493L..21L,2010Sci...329...57L}. Besides this planet,
only a couple of planetary-mass objects have been detected around
stars with a debris disk. In fact, most debris disks have been imaged
around young bright A- or F-type stars, which are either rapid rotators
and/or active stars. The search for companions with radial velocity
techniques around these stars is therefore extremely difficult. Key
breakthroughs have recently been made with high-contrast imaging: a
$<3$-\Mjup\ planetary companion was detected in the outskirts of
Fomalhaut's debris disk \citep[119~\au\ from the
  star;][]{2008Sci...322.1345K}, while four planetary companions of 7,
10, 10, and 10~\Mjup\ were imaged at 68, 38, 24, and 14~\au , respectively, from HR~8799
\citep{2008Sci...322.1348M,2010Natur.468.1080M}. It is not clear
whether these objects could have formed through core accretion -- and not from gravitational instabilities -- the
foremost theory of planet formation, although \betapic~b is a serious
candidate for this formation scenario \citep{2010Sci...329...57L}.

Most companions detected with high-contrast imaging, such as
AB~Pic~b \citep{2005A&A...438L..29C} or 2M1207~b 
\citep{2004A&A...425L..29C}, are believed to be either too far or 
too massive to have formed through core accretion. An alternate pathway
gaining widespread interest is a stellar formation mechanism, 
i.e., formation during the fragmentation and collapse of a molecular cloud. 
There are many unanswered questions raised by these models. 
Do these two processes for forming planets and brown dwarfs/low-mass stars 
operate exclusively (does one inhibit the other)? Can core accretion from
a disk operate when a brown-dwarf companion exists at large separation? 
In contrast, can disks and planets exist around tight binary systems,
composed of the primary star and a very close low-mass star or brown
dwarf? How do massive companions impact the dynamic evolution of
planets and disks? 

To answer these questions, it is necessary to probe the innermost
  regions of debris disks when searching for substellar companions. For
  the closest debris disk, direct imaging proved to be a successful
  technique, as shown for $\beta$~Pic or Fomalhaut. But most of
  the young associations (as well as stellar forming regions) lie at
  more than a hundred parsecs. At such a distance, 10 \au\ is equivalent
  to 100\,mas, less than twice the diffraction limit of an 8-meter
  telescope in the near-infrared. This observational domain is not
  filled by differential imaging techniques, whose inner working angle
   is typically of few resolution elements.  It is neither
  filled by long baseline interferometry: the field of view of a
  single mode interferometer is too small, equal to the resolution
  element of one of its telescopes. The effectiveness of both
  techniques is reviewed in detail in
  \citet{2010A&ARv..18..317A}. In terms of angular separation,
  aperture masking is the missing link in the field of high dynamic
  observations. It gives a relatively high dynamic range (5 to 7 mag)
  at a specific angular separation ($\approx 100\,$mas) which is of
  greatest interest for planetary formation.

   In this paper, we present three stars observed with an aperture
   mask at the VLT.  In the next section, the instrument is described,
   along with the observational procedure and data reduction methods. 
   In section~\ref{test}, we present a clearly resolved detection of the binary star
   HD\,135549 for illustrative and pedagogical 
   purposes. Section~\ref{debris} reports on the non detection contrast
   limits obtained with SAM for the two debris disks HD\,92945 and
   HD\,141569. We present these limits as a function of flux ratio and 
   infer the mass limit derived from current models.


\section{Observation \& data reduction}
\label{reduc}

\subsection{SAM at NaCo}

An aperture mask works by transforming the telescope into a Fizeau
interferometer. The point spread function is a complex superposition
of fringes at given frequencies.  In specific cases, pupil masking can
outperform more traditional differential imaging for a number of
reasons.  First, the masks are designed to have nonredundant array
configurations that permit phase deconvolution; slowly moving optical
aberrations not corrected by the AO can be accuratly
calibrated. Second, the mask primarily rejects baselines with low
spatial frequency and passes proportionately far more baselines with
higher $\lambda/B$ resolution than does an orthodox fully filled
pupil.  Third, high-fidelity recovery of phase information allows
``super resolution'', with a marginal loss of dynamic range up to
$\lambda$/2D.  The principal drawback is a loss in throughput so that
photon and detector noise can affect the signal-to-noise ratio even
where targets are reasonably bright for the AO system to close the
loop.  In common with all synthesis imaging by interferometry, the
effective field-of-view of SAM is determined by the shortest baseline
so that the technique is not competitive at separations larger that
are greater than several times the formal diffraction limit.

VLT's SAM mode has grown from earlier seeing-limited experiments on
the Keck telescope \citep{2000PASP..112..555T}, and more recently has
gained widespread use on adaptive optics equipped telescopes
\citep{2006SPIE.6272E.103T, 2006ApJ...650L.131L}.  Commissioning of
SAM at NaCo has required important software development from ESO at
Paranal. It prompted the development of datacube storage (offered now regularly on NaCo as ``cube mode'')
 and of pupil-tracking observation (the Nasmyth
rotator tracks the telescope elevation). SAM is currently under
investigation as an additional mode for the VISIR instrument,
operating in the mid-infrared. It is under informal study for the
SPHERE instrument \citep{2008SPIE.7014E..41B}.

\subsection{Observing procedure}
\label{procedure}

\begin{table*}
\caption{Observation log}          
\label{log}    
\centering     
\begin{tabular}{lccccccc} 
\hline\hline             
Target & Date & Start time (UT)& End time (UT) &  DIT (ms) & NDIT & Dithers & Acquisitions\\
\hline                      
 HD\,135549   & 2010-03-18 & 05:43:27 & 10:13:37 & 0.2 & 200 & 8 & 15 \\ 
 HD\,135344b  & 2010-03-18 & 05:32:31 & 10:04:16 & 0.2 & 200 & 8 & 15 \\   
\hline                      
 HD\,92945  & 2010-04-07 & 02:06:41 & 03:01:27 & 0.4 & 100 & 8 & 4 \\                   
 HD\,92933  & 2010-04-07 & 02:14:49 & 02:52:55 & 0.4 & 100 & 8 & 3  \\   
\hline                      
 HD\,141569  & 2010-04-07 & 07:16:35 & 08:33:03 & 1.2 & 35 & 8 & 4 \\   
 BD-03\,3826 & 2010-04-07 & 07:34:55 & 08:24:37 & 1.2 & 35 & 8 & 3 \\   
\hline                                   
\end{tabular}
\end{table*}

Observations were taken in the L prime band ($\lambda=3.80\pm0.31
\mu$m) using the ``7~holes'' mask and the IR wavefront sensor
(WFS). The diffraction pattern is plotted in Fig.~\ref{Image}.  The
physical properties of the mask -- as well as the three others --
are described in the NaCo
manual\footnote{http://www.eso.org/sci/facilities/paranal/instruments/naco}.
The complete observational sequence for an object typically took two
hours.  Table~\ref{log} contains observing details such as integration
times, datacube size, and number of acquisitions.

Each science target is associated to a PSF reference star situated
within one degree. 
A single observing block (OB) is used that encodes replicated observations for
both calibrator and target.  Rapid star acquisition -- which
is important for ensuring good calibration -- can be augmented by using a
simple offset of the telescope without having to preset the telescope
and perform a full (instrumental) re-acquisition.  In this mode, which
has been christened ``star hopping'', the adaptive optics loop is
opened, while the template orders a dither to bring a different star
into the AO field selector. The AO is then closed manually by the
operator without incurring the time penalty for re-optimization.
Star hopping therefore only works on objects of comparable brightness
in the wavefront sensor.

As the observation progresses, the repetition of the template collects 
eight datacubes of multiple frames (typically a hundred), each at a 
different dither position on the detector. For these observations, the
detector was windowed to $512\times 514$ pixels.

\begin{figure}
   \centering
   \includegraphics[scale=.3]{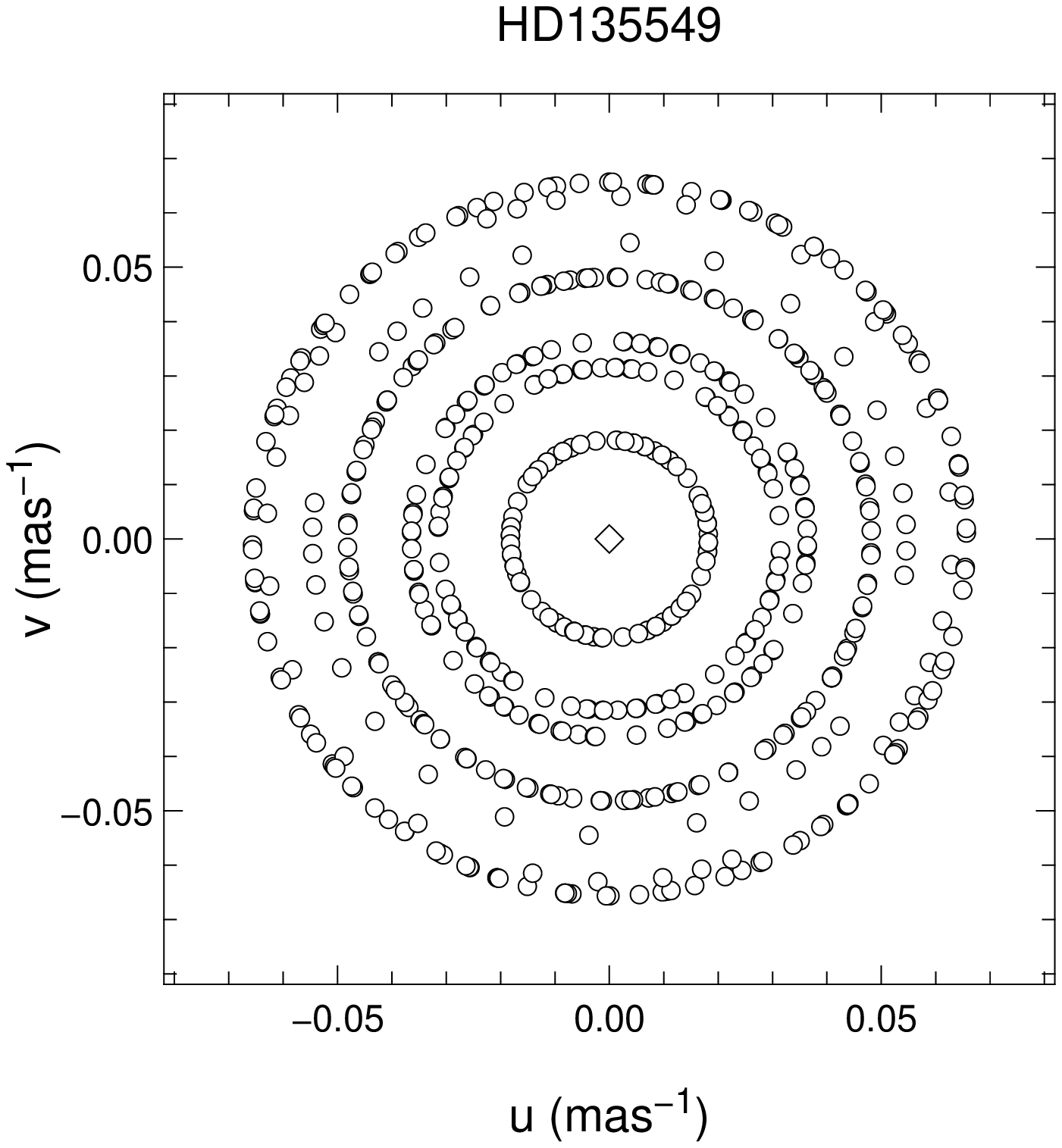}\\
   \includegraphics[scale=.3]{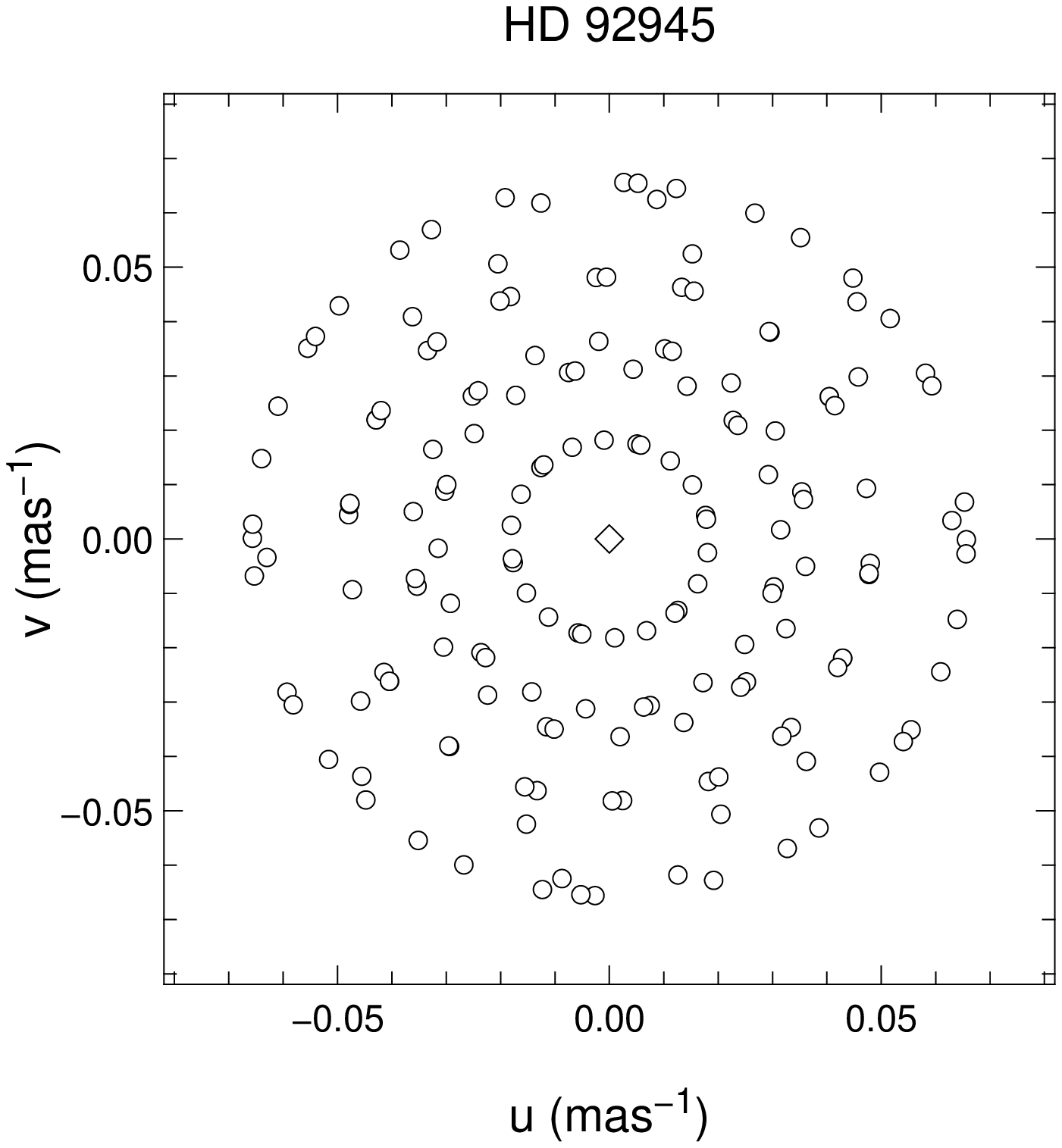}
   \includegraphics[scale=.3]{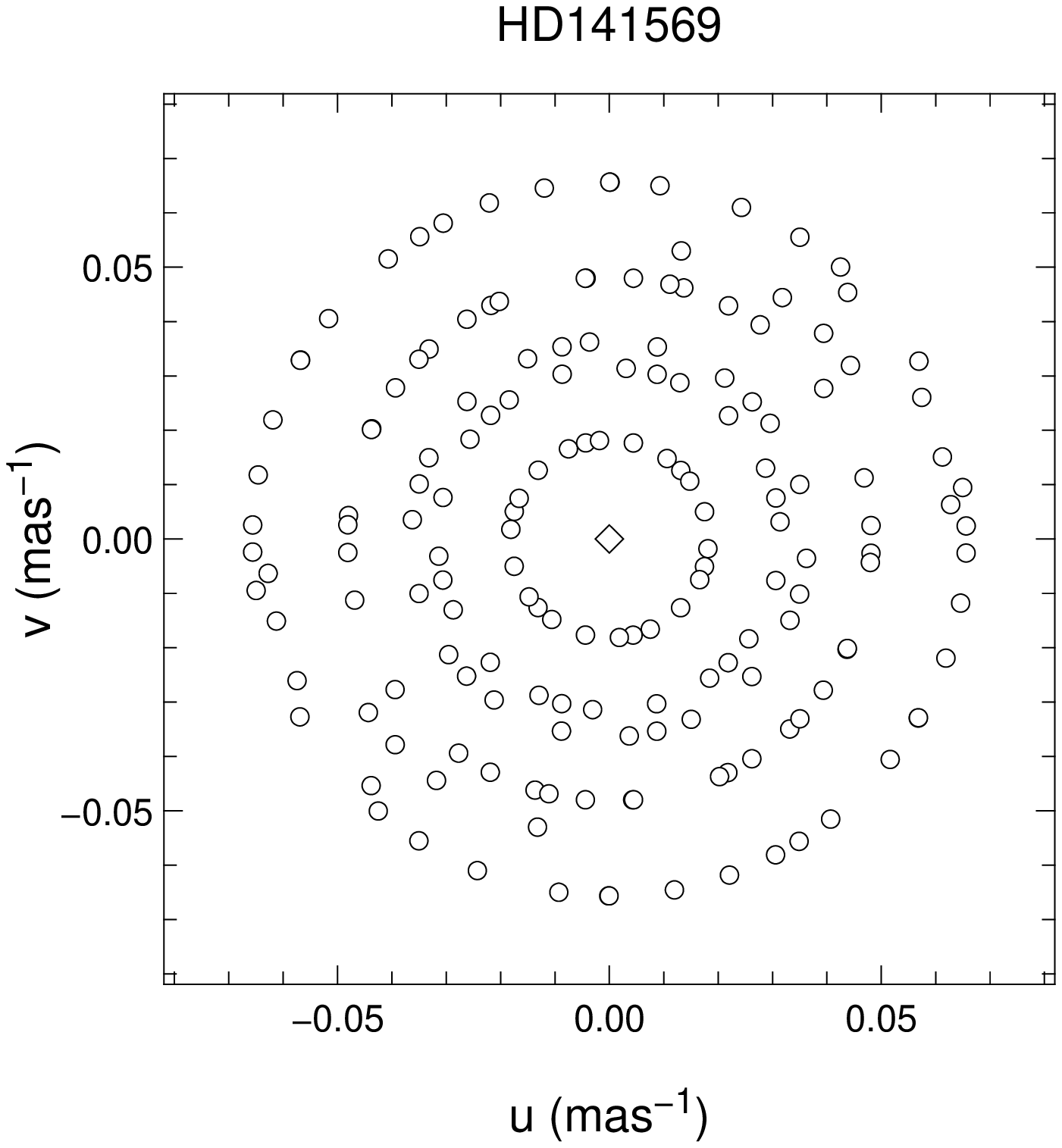}
   \caption{Frequency coverage for targets HD\,135549, HD\,92945, and
     HD\,141569.}
              \label{uv}%
    \end{figure}

A single snapshot observation yields Fourier coverage of 21 spatial
frequencies. However, the pupil-tracking mode results in sky rotation 
on the detector as the parallactic angle changes. This variation with time
results sweeps the baselines into circular Fourier tracks and permits
rotational aperture synthesis techniques to assist with the filling of 
the spatial frequency UV-plane plane, as illustrated in Fig.~\ref{uv}.

\subsection{Data reduction}

   \begin{figure}
   \centering
   \includegraphics[scale=.45]{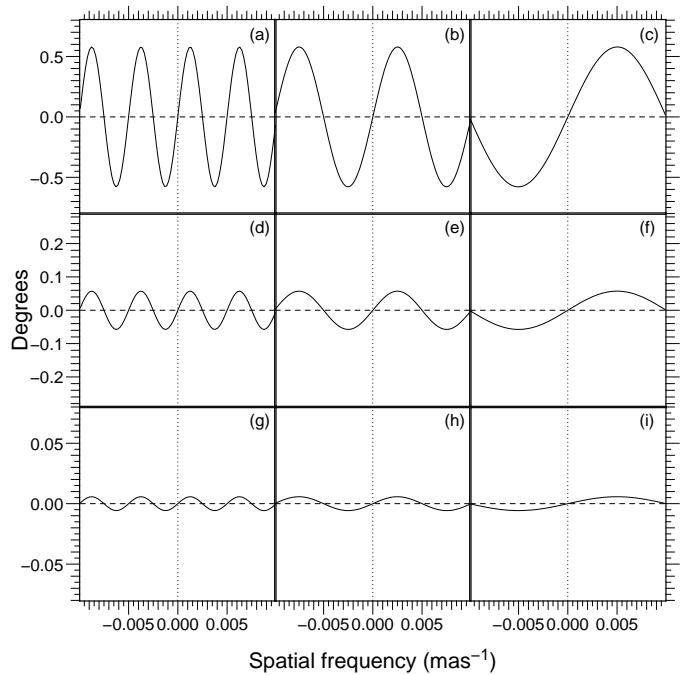}
   \caption{Phase as a function of the spatial frequency. The
     different plots are models of a binary system with different flux
     ratio and angular separation.  No noise has been added, so
       the curves represent the perfect case. From top to bottom, the
     contrast of the primary to secondary decrease. Panels (a-b-c):
     flux ratio of 1\%. Panels (d-e-f): flux ratio of 0.1\%. Panels
     (g-h-i): flux ratio of 0.01\%. From left to right, the angular
     separation gets smaller.  Panels (a-d-g): 200\,mas.  Panels
     (b-e-h): 100\,mas.  Panels (c-f-i): 50\,mas. In the L' band, the
     cut-off frequency of an eight meter telescope is
     0.01\,mas$^{-1}$.}
              \label{Model}%
    \end{figure}

   \begin{figure}
   \centering
   \includegraphics[scale=.5]{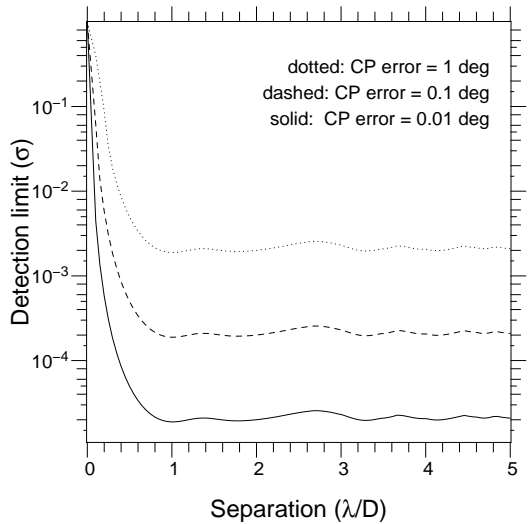}
      \caption{Theoretical (Monte-Carlo) 1\,$\sigma$ detection limits of the 
        7-hole mask as a function of angular separation and error on
        closure phases. The ripples on the detection limit curves are
        not due to uncertainty in the Monte-Carlo simulation, but to
        the nonhomogeneous coverage of the UV frequency plane.
      }
         \label{Monte}
   \end{figure}

To the author's knowledge, two data reduction software libraries are
presently available\footnote{upon request to developers} to reduce SAM
data. One pipeline has been developed by Sydney University, Cornell
University, and Caltech from 2004 onwards, based on an earlier
pipeline from Berkeley. It has already been used for several papers
arising predominantly from Keck aperture masking data
\citep{2000PASP..112..555T}. This reduction algorithm is based on fast
Fourier transform (FFT).  The data presented in this paper have been
reduced by the sparse aperture mode pipeline (SAMP) developed at the
Observatoire de Paris. This software is similar to that from Sydney
university, and numerous tests and cross-checks have produced similar
results.

For both pipelines, data reduction follows a similar path:
\begin{itemize}
\item {\bf Pre-reduction}: the data are (i) flatfielded, (ii) bad-pixel
  corrected, and (iii) background-subtracted.
\item {\bf Complex visibility}: Complex visibilities are obtained from
  FFT of the image (Sydney pipeline), or by direct fringe fitting (SAMP)\footnote{ Fringe fitting -- in a least squares sense -- is performed through matrix inversion thanks to singular value decomposition}.
\item {\bf Bispectrum}: the bispectrum is obtained using
  multiplication of the complex visibilities corresponding to a
  triangle of holes in the mask. Bispectrum is averaged over all the
  frames of each set of eight datacubes. Closure phases are obtained
  using the argument of the bispectrum.
\item {\bf Calibration}: Closure phases obtained on the calibrator are
  fitted with a polynomial function, and subtracted from the closure
  phases on the science object.
\item {\bf Detection}: The closure phases are fitted with a model of
  binary star (see section~\ref{test}). A $\chi^2$ map is computed,
  giving the likelihood of the presence of the companion for any given
  separation and position angle.
\end{itemize}

Atmospheric dispersion is not accounted for in our pipeline, but
  we are aware that in certain conditions (presence of emission lines
  in the spectrum), it can mimic  asymmetry along the elevation
  axis. Such false detection can be avoided by observation at
  different parallactic angles.

The effect of fringe blurring due to the large bandwith of L' filter
can technically be accounted for in the fitting process. It would
require including a loss of fringe coherence on the edge of the
diffraction pattern. However, at this stage, the SAMP pipeline does
not include this correction. It is, however, not seen as critical with
the seven-hole mask, since the diffraction pattern of the 1.2 meter holes
are significantly smaller than the coherence length (see
Fig.~\ref{Image}). It becomes an acute problem with the small holes of
the 18-hole mask, which then justifies using a narrow band filter.

The complex visibilities consist of amplitudes and phases, but for the
purposes of binary detection, we typically rely on the phases only
(of the bispectrum). This is because of the dependence of
absolute visibilities on the adaptive optics correction. They
are thus harder to calibrate, and are subject to different biases between the
science and the calibrator object.  However, by default, the pipelines
also process the power spectrum and estimate fringe
visibilities. Performance of SAM's visibility estimation will be the
subject of a forthcoming paper.

\section{Test case: the HD\,135549 binary system}
\label{test}

\subsection{Principles of faint companion detection with SAM}
\label{princ}

Aperture masking data yields Fourier phases and amplitudes. They are
functionally equivalent to those recovered in long-baseline
interferometry, and therefore the same techniques and software can be
applied. Image reconstruction is a possibility, but to
fulfill the difficult requirements of faint companion detection,
direct fitting in the Fourier domain is essential.

Modeling the presence of a two point-like sources in the Fourier domain is
trivial. The complex visibility can be written as:
\begin{equation}
V(u,v)=1-r\,\left\{1-\exp \left[i 2 \pi (-\alpha u + \delta v)/\lambda\right]\right\} \, ,
\label{eq}
\end{equation}
where $u$ and $v$ are orthogonal spatial frequency vectors (respectively 
to the east and to the north), $\alpha$ and $\delta$ are the right 
ascension and declination in radians, and $r$ is the flux ratio between 
the primary and secondary. Figure~\ref{Model} presents the interferometer 
fringe phase for a grid of nine models with three different angular 
separations and three different flux ratios. A direct relationship
between the phases and the basic properties of the binary system can
be noted. The phases exhibit a sinusoidal curve across the Fourier
plane, with the amplitude proportional to the contrast ratio while
the period is proportional to the separation.
Closure phases are derived from the baseline phases as a  
simple summation of the three phases corresponding to a closed
triangle of holes.
 
Companion searches are performed by fitting the various binary
parameters ($r$, $\alpha$, and $\delta$) to the measured closure
phases, so the detection limit obtained from any given dataset is
directly related to the signal-to-noise ratio of the closure phase. A
Monte-Carlo simulation was performed to test the theoretical
detection limit, as a function of the angular separation and closure
phase measurement error bars. The result is presented in
Fig.~\ref{Monte}. The detection limit is almost constant for any
separation larger than $\lambda$/D. The Monte-Carlo simulation shows that the dynamic range can be
approximated as (as a function of $\sigma(CP)$ in degrees)
\begin{equation}
detection\ (1\sigma)= 2.5\times10^{-3} \times \sigma(CP) \, .
\label{dyn}
\end{equation}
Detections below $\lambda$/D can still be significant, but the
contrast achieved decreases rapidly with the separation.  At half
$\lambda$/D, limits are worse by a factor of two, while at
0.25\,$\lambda$/D the degradation in performance is a factor 15. For separations beyond the formal diffraction limit, any fitting
process becomes highly model dependent and often strong degeneracies
between fitted parameters can appear.  A typical example at small
separations is the covariance between the flux ratio and separation: it
becomes difficult to measure these two parameters independently.

\subsection{Detection of the stellar companion to HD\,135549}

HD\,135549 was initially observed for the purpose of a PSF reference
star, but it turned out to be a binary system with a contrast ratio
around 5\% and a separation close to 180 mas. We present the data in
this paper for the sake of introducing the technique of detecting
faint companions with SAM illustrated by strong, clear systematic
signals.

\begin{table*}
\caption{\label{result_bin} Detection as a function of time using
  individual pairs of source-calibrator files. }
\centering
\begin{tabular}{lcccc}
\hline\hline
UT & Position Angle & Separation& Position Angle& Flux \\
 & on detector (deg) & (mas) & on sky (deg) & (\% of central object) \\
\hline
05:43:27& $-153.38$ & $179.32 \pm 0.38$  & $148.24 \pm 0.13$  & $5.62 \pm 0.07$   \\ 
06:03:35& $-148.63$ & $179.33 \pm 0.37$  & $147.96 \pm 0.11$  & $5.51 \pm 0.07$   \\ 
06:21:57& $-143.54$ & $181.52 \pm 0.26$  & $148.25 \pm 0.09$  & $5.69 \pm 0.04$   \\ 
06:40:14& $-137.46$ & $179.55 \pm 0.58$  & $148.64 \pm 0.21$  & $5.60 \pm 0.07$   \\ 
06:58:02& $-130.17$ & $179.46 \pm 0.37$  & $148.28 \pm 0.10$  & $5.62 \pm 0.06$   \\ 
07:16:34& $-120.66$ & $179.24 \pm 0.30$  & $149.15 \pm 0.12$  & $5.60 \pm 0.05$   \\ 
07:35:08& $-108.48$ & $179.42 \pm 0.29$  & $148.41 \pm 0.10$  & $5.57 \pm 0.06$   \\ 
07:53:06& $-93.93$ & $179.34 \pm 0.46$  & $148.59 \pm 0.18$  & $5.45 \pm 0.08$   \\ 
08:11:23& $-77.02$ & $180.50 \pm 0.91$  & $148.10 \pm 0.21$  & $5.43 \pm 0.08$   \\ 
08:29:18& $-60.32$ & $180.80 \pm 0.50$  & $148.41 \pm 0.19$  & $5.47 \pm 0.07$   \\ 
08:55:48& $-39.29$ & $181.44 \pm 0.28$  & $148.72 \pm 0.10$  & $5.58 \pm 0.04$   \\ 
09:13:54& $-28.29$ & $180.80 \pm 0.30$  & $148.50 \pm 0.12$  & $5.57 \pm 0.05$   \\ 
09:37:18& $-17.41$ & $181.60 \pm 0.39$  & $148.56 \pm 0.11$  & $5.27 \pm 0.07$   \\ 
09:55:46& $-10.75$ & $180.40 \pm 0.54$  & $148.07 \pm 0.16$  & $5.33 \pm 0.07$   \\ 
10:13:37& $-5.44$ & $179.73 \pm 0.34$  & $147.61 \pm 0.13$  & $5.49 \pm 0.04$   \\ 
 \hline
Full Fit& {\em all} & $180.02 \pm 0.11$  & $148.36 \pm 0.04$  & $5.53 \pm 0.02$  \\ 
 \hline
\end{tabular}
\end{table*}

\begin{figure}
\centering
\includegraphics[scale=.37]{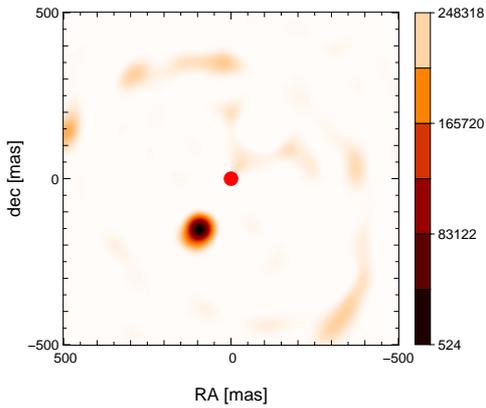}\\ 
   \caption{$\chi^2$ surface as a function of $\alpha$ and $\delta$ obtained
     by fixing $r$ to the best-fit value in the model of a binary stated in
     Equation~\ref{eq}. A clear minimum indicates the position of the
     stellar companion.}
              \label{plot_map}%
    \end{figure}

\begin{figure*}
\centering
   \includegraphics[scale=.37]{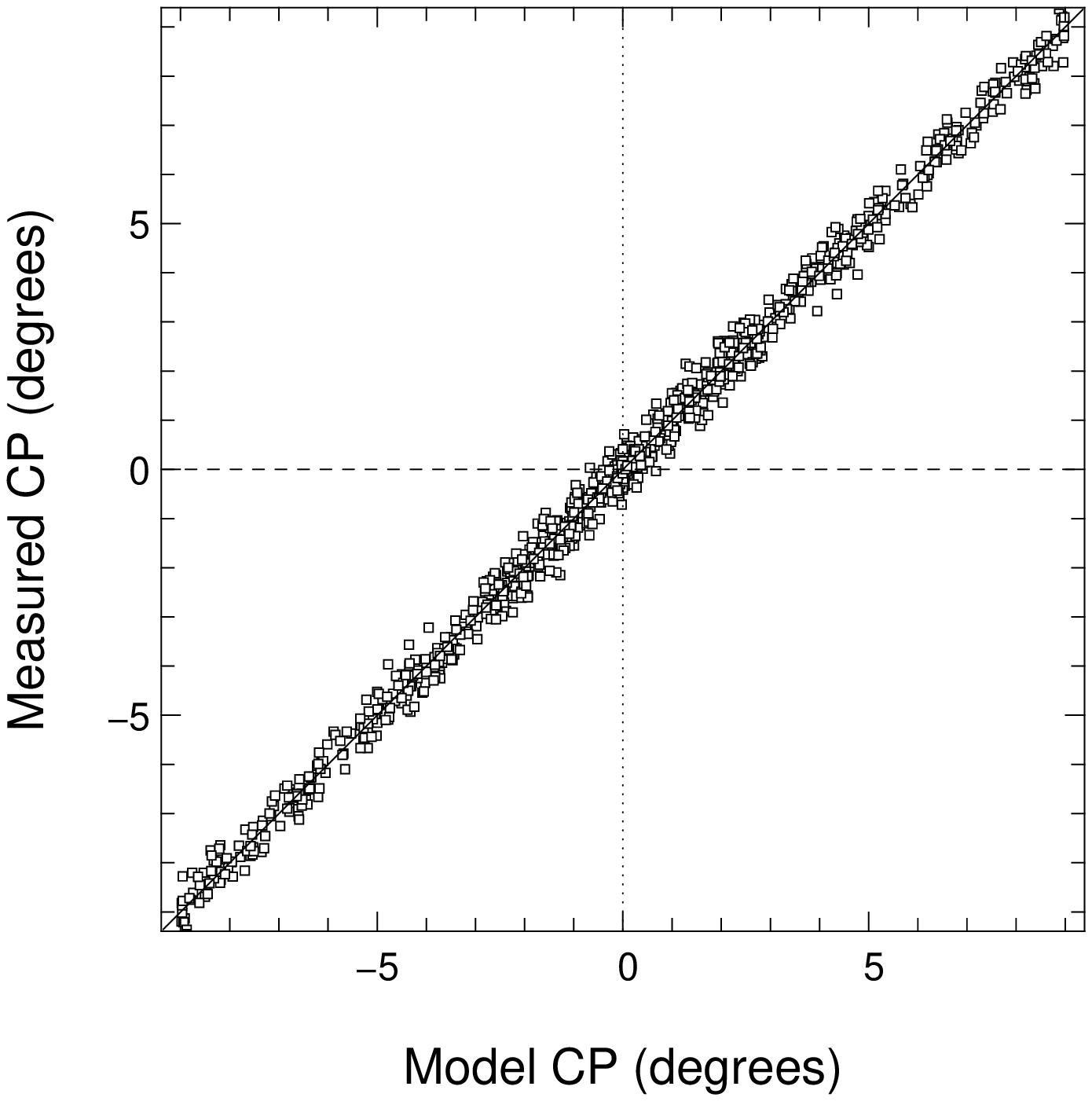}\hspace{1cm}
   \includegraphics[scale=.37]{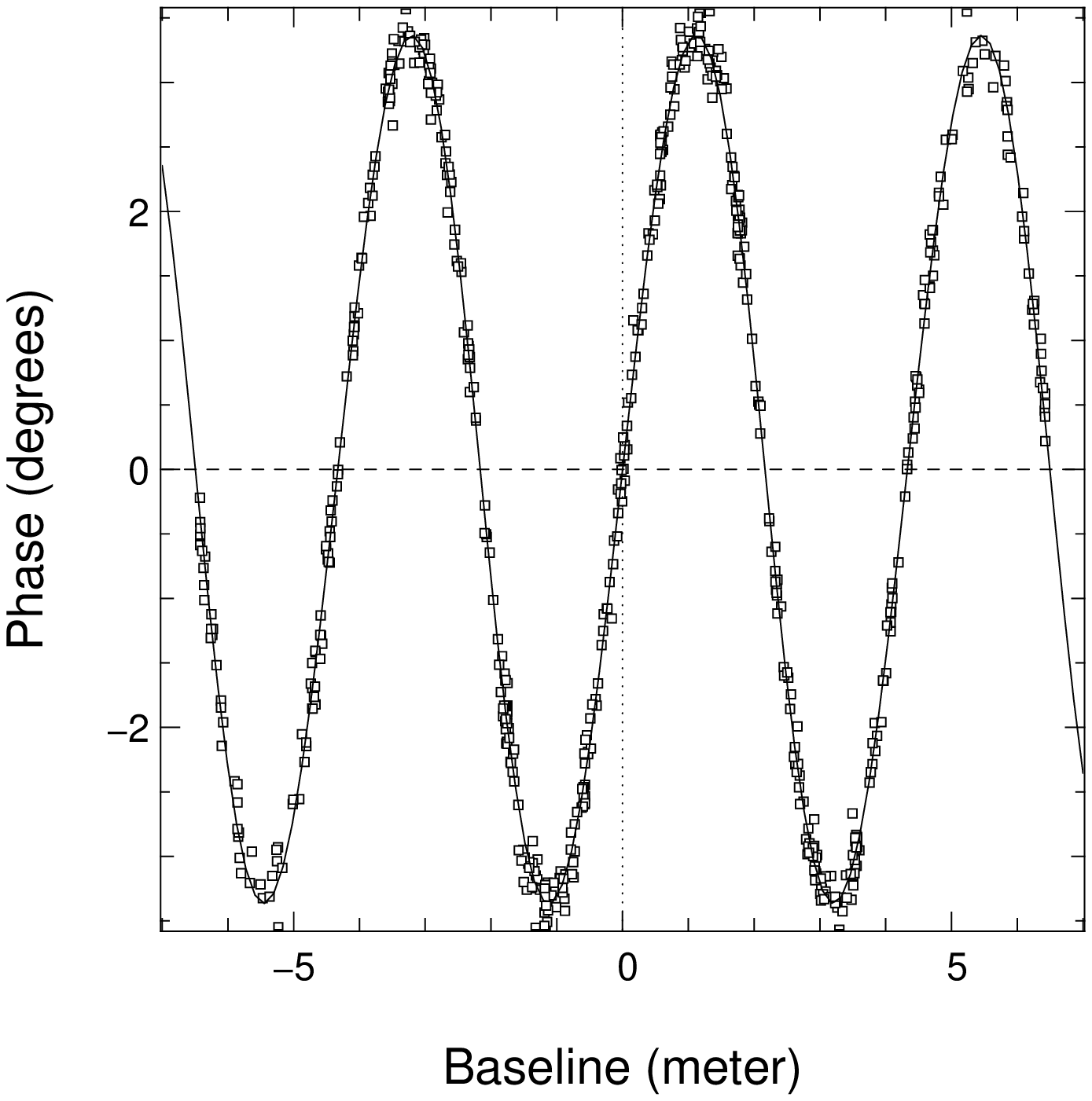}
   \caption{Lefthand panel: Measured closure phase as a function of
     model closure phase assuming the best-fit binary parameters. The
     rms error between the model and the measurement is 0.28\,degrees.
     Righthand panel: phase obtained by least square inversion of the
     closure phase to phase matrix, plotted as a function of the
     projected baseline (perpendicular to the binary orientation;
     hence the sinusoidal curve).}
              \label{plot_res}%
    \end{figure*}

   \begin{figure*}
   \centering
   \includegraphics[scale=.88]{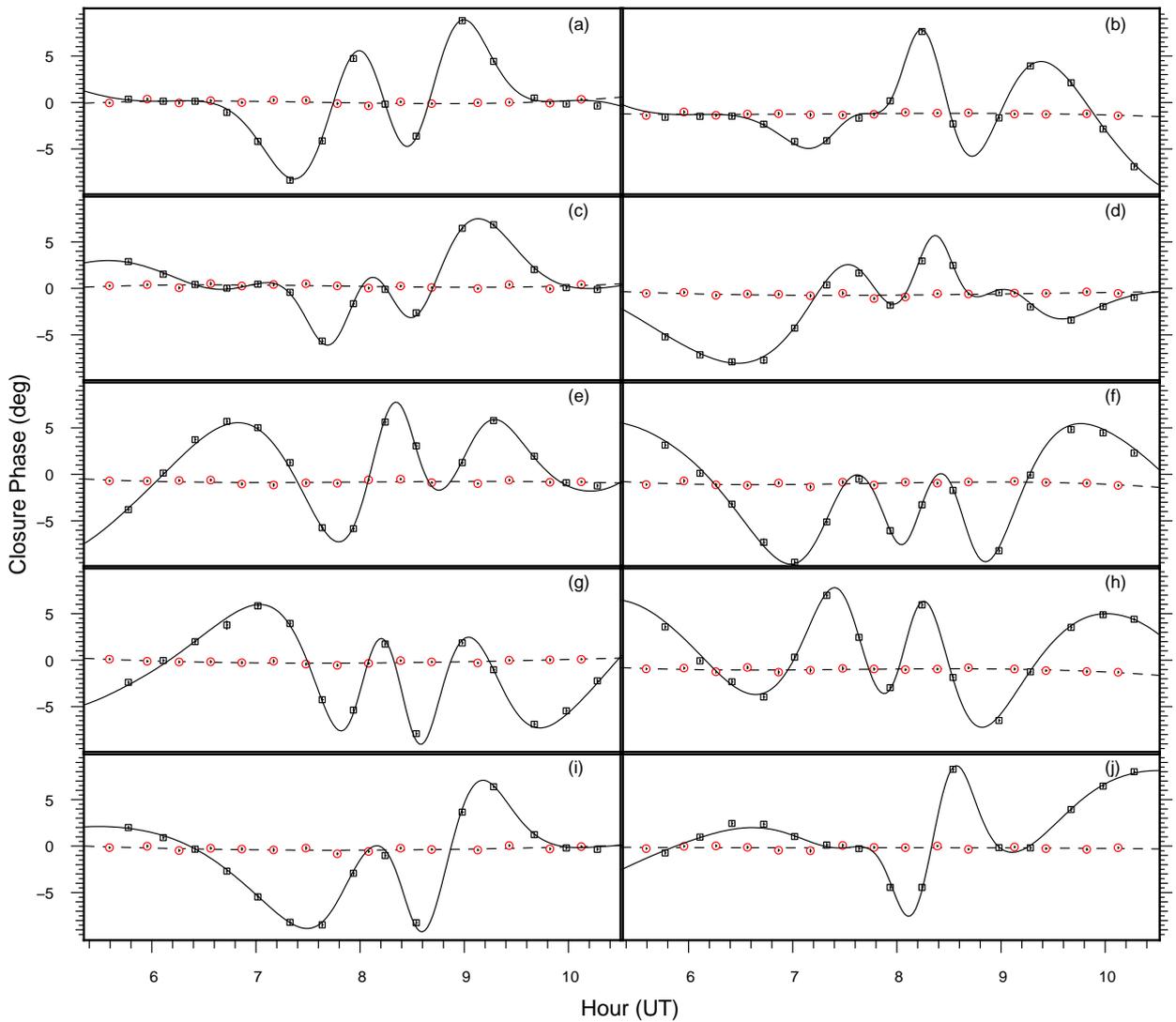}
   \caption{Closure phase as a function of time for the 10 largest
     triangles. The triangles are made of the holes (a) 1-5-7, (b)
     1-2-6, (c) 1-6-7, (d) 4-5-6, (e) 1-2-7, (f) 1-5-6, (g) 4-5-7, (h)
     1-4-7, (i) 2-5-7 and (j) 2-5-6; numbered according to the rows of
     Table 4-19 in the NaCo manual (available online at eso.org).
     Calibrator data are represented as circles, while binary star
     data are squares. The dashed line is the polynomial fit of the
     calibrator, and the solid line is the closure phase predicted by
     the best fitting binary system model. The same fit quality is
     obtained on all 35 closure phases. Overall, the average rms error
     of the fit is 0.28 degrees.}
              \label{fit}%
    \end{figure*}

 The free parameters of Eq.~(\ref{eq}) are the binary coordinates
 $\alpha$ and $\delta$, along with the flux ratio $r$.  A map of the
 reduced $\chi^2$ is derived as a function of $\alpha$ and $\delta$
 (Fig.~\ref{plot_map}). It indicates the presence of a companion with
 a strong likelihood at a separation of $\approx 180$\,mas to the
 southeast of the star. The best-fit model is compared to the observed
 data in the two panels of Fig.~\ref{plot_res}.  The lefthand panel
 shows the observed versus model-predicted closure phase data. The
 final reduced $\chi^2$ is equal to 1.85 while the average residual is
 0.28 degree over a total of $35\times15=525$ closure phase
 measurements.  According to Eq.~(\ref{dyn}), this dataset therefore
 has the capability of a detection limit of $7.5\times10^{-4}$
 (1\,$\sigma$ detection).  The righthand panel shows phases recovered
 from the closure phases. This is done by inverting the phase to
 closure phase matrix. The matrix is not fully invertible; for
 example, tip-tilt is not constrained by the closure phase. Therefore,
 unknown parameters of the inversion were taken to match the
 model. The best fit model is over-plotted as a solid line.

The first 15 lines of Table~\ref{result_bin} results from the
processing of each SCI-CAL acquisition pair separately. An independent
model is fitted at each epoch, and the detection can be seen to be rotating 
in the detector frame (but fixed on the sky). A strong indication that our 
detection is indeed real lies in the rotation in the detector
frame following the parallactic angle evolution. Because the telescope
is azimuthal, it is unlikely that a calibration error (including
optical aberrations) rotate in such a motion. This
validation could not have been made with an equatorial telescope.

The last line of Table~\ref{result_bin} gives the best-fitting binary
parameters. Separation and dynamic ranges are obtained with high
statistical precision. The precision of the closure phase measurement
over time is highlighted in Fig.~\ref{fit}, in which each panel
depicts a different closure phase. The solid line corresponding to the
model can be seen to almost perfectly match the evolution of the
closure phase as a function of time.

\subsection{On the limitation of the dynamic range}

As stated by Eq.~\ref{dyn}, the dynamic range is a pure function of the error on the phase.
The binary system was observed during good atmospheric conditions
(0.7'' seeing and a coherent time of 8\,ms) with a relatively fast frame time of
200\,ms.  Phases recovered from the interferograms show a median
frame-to-frame error of 17 degrees, equivalent to a 180-nm piston
jitter at our observing wavelength.  On the other hand, the
frame-to-frame closure phase error is only 2.8~degrees. This is a
clear indication that most of the 180-nm rms error is due to the
residual of atmospheric perturbation not corrected by the AO system.

Assuming that these residuals were caused by photon noise or
readout noise, the remaining 2.8~degrees frame-to-frame error should
be uncorrelated. In this case, the final error on a whole
acquisition sequence (8 dither of 200 frames) should improve as the
square root of the number of frames giving an expected error of  
0.07~degrees. For such a scenario, the dynamic range obtained in the 
final detection should be $1.7 \times 10^{-4}$ (according to
Eq.~(\ref{dyn})). Unfortunately, the observed standard deviation for both
calibrator and the science target closure phases are significantly 
larger (0.25 and 0.33~degrees respectively; obtained by comparing
different telescope dithers). These larger errors point to an additional
systematic term generating biases in the phase closure which do not
average out over long data runs. In order to extract the very best
contrast ratio in our faint companion detection strategy, these
closure phase biases need to be understood and calibrated.

Within our present software pipeline, closure phase calibration is
performed by subtraction of the PSF reference star closure phases which resulted
in an rms closure phase error observed of 0.24~degrees (when comparing
different dither positions).  Compared to the theoretical limit of the
high-frequency noise ($\sqrt{2}\times 0.07 \approx0.1$\,degrees), it
is apparent that systematic biases in the closure phase remain the
dominant limit to the dynamic range.  When fitting the final data with
a model of a binary system, the
residual error on the closure phases is 0.28~degrees.

 This statistical analysis shows that the high-frequency noise --
  assumed to be photon and readout noise -- is not limiting the
  dynamic range. On this specific dataset, one magnitude in contrast
  ratio could be gained if only it was possible to have a perfect
  calibration of the closure phase systematics. A higher dynamic range
  could then be achieve by repetition of the same
  observation. The reason for these systematics errors is still under
  debate. Between several possibilities, we can mention i) optical
  aberrations before the pupil mask, ii) correlated detector read out
  noise, iii) spatially correlated phase error due to the modal
  correction of the adaptive optic system.  The residuals could
   be diminished partially by observing several calibrators. But not to
  a full extent: the noise would average only on the calibrator
  dataset.

\section{The two debris disks HD\,92945 and HD\,141569}\label{debris}

   \begin{figure*}
   \centering
   \includegraphics[scale=.37]{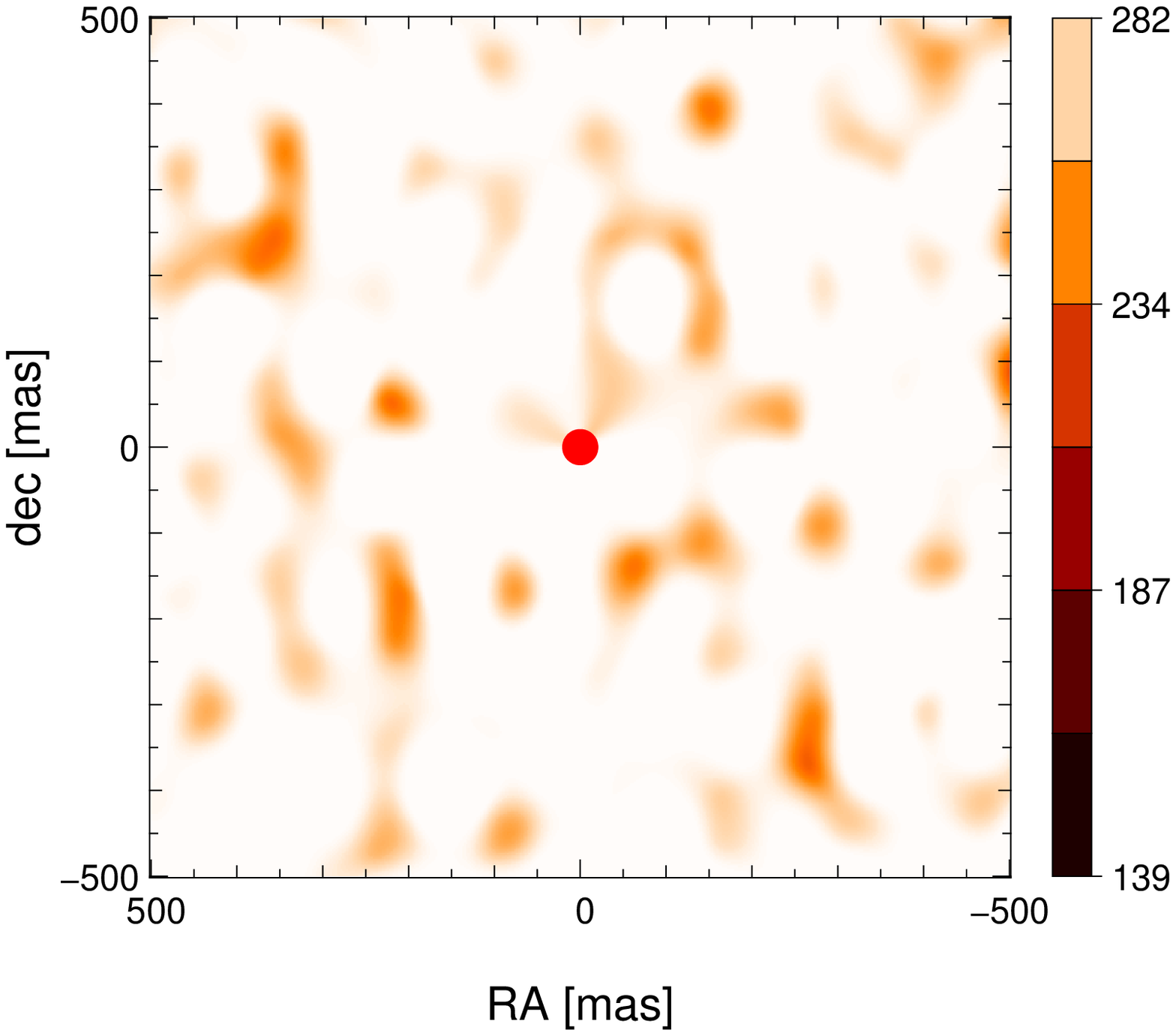}\hspace{1cm}
   \includegraphics[scale=.37]{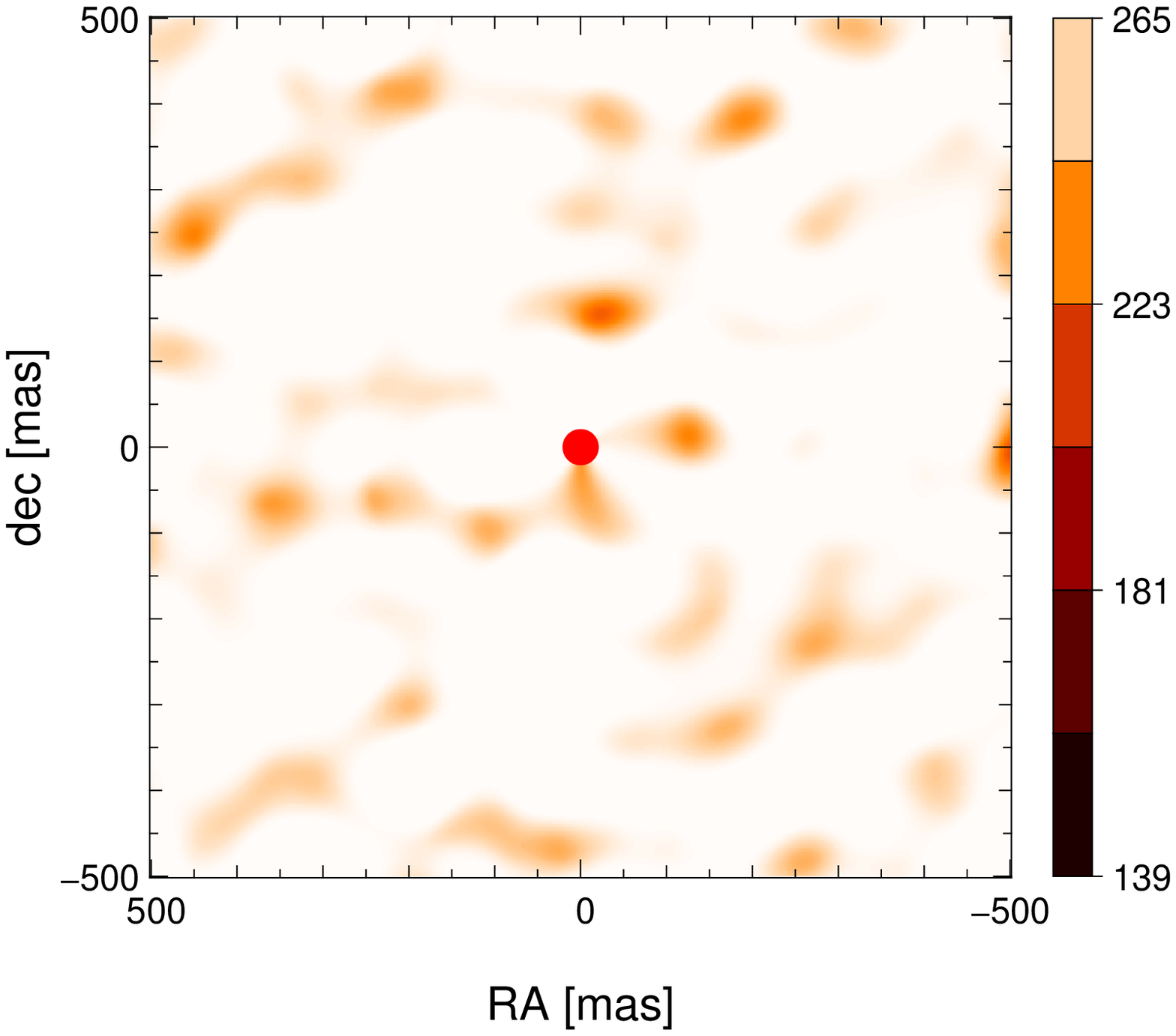}
   \caption{$\chi^2$ as a function of $\alpha$ and $\delta$ obtained
     from the best fit of $r$ in the model of a binary stated in
     Equation~\ref{eq}. Left panel: HD\,92945. Right panel:
     HD\,141569. The lower limit of the color scale (139) is the
     number of degree of freedom of the fit.}
              \label{chi2_map_nd}%
    \end{figure*}

\subsection{The targets}
\label{targets}

\subsubsection{HD\,92945}

HD\,92945 (V419\,Hya) is a K1.5V star located at 22~pc.
  \citep{2004ApJ...614L.125S} estimated an age of 100~Myr using
  \ion{Ca}{ii} lines.  If we suppose HD\,92945 to be part of the AB
  Dor group \citep{2006ApJ...643.1160L}, this age agrees with
  the estimation of $\approx 50$-120~Myr for the whole group
  \citep{2004ApJ...613L..65Z,2005ApJ...628L..69L}. Observations with
the Advanced Camera for Surveys (ACS) on the \emph{Hubble Space
  Telescope} have revealed it is surrounded by a disk $\sim30\degr$
from edge-on with a 0\farcs7-thick bright ring at 57~\au\ from the
star. The disk has a diffuse component detected between 55 and
170~\au\ \citep{2007lyot.confE..46G}. The inner gap may well be
  the result of resonances with a stellar or planetary companion
  \citep{1994ApJ...421..651A}.

\subsubsection{HD\,141569}

HD\,141569 is a $5\pm3$-Myr-old \citep{2004A&A...419..301M} pre-main
sequence B9.5V star in a triple system. It is located 99-pc away and
the central star features an elegant debris disk that was first imaged in the
near-infrared \citep{1999A&A...350L..51A} and in visible scattered
light \citep{2003AJ....126..385C}. The two other components of the
system form an M2-M4 binary, $\sim7\farcs5$ away from the central
star. \citet{2009A&A...493..661R} give an overview of the rich disk 
structure. It is composed of two annuli with peak luminosity in
scattered light at $\sim200$ and $\sim325$~\au\ from the star. These
two rings have azimuthal brightness asymmetries (factors 2 to 3). The
outer ring shows a tightly wound spiral structure. The rings are
separated by an apparent gap wider than the rings. An extended diffuse
emission associated with a faint spiral arm is present in the
northeast side of the disk, and is detected to more than
600~\au. Another spiral arm is possibly observed, pointing toward the
binary companions. Finally, the disk brightness sharply decreases
within 200~\au, suggesting a strong dust depletion in the innermost
regions of the disk.

\subsection{5\,$\sigma$ detection limits}

   \begin{figure*}
   \centering
   \includegraphics[scale=.37]{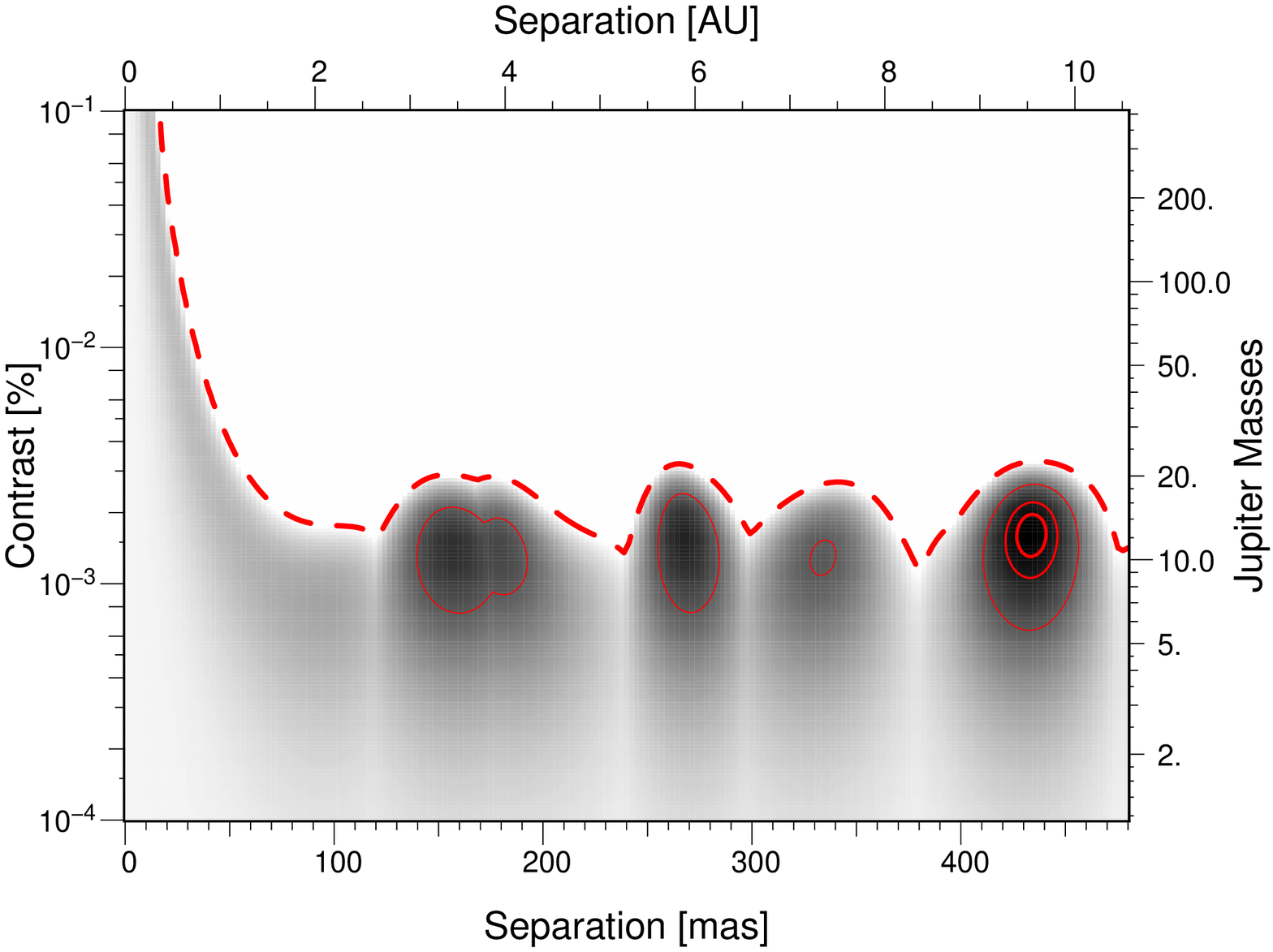}\hspace{1cm}
   \includegraphics[scale=.37]{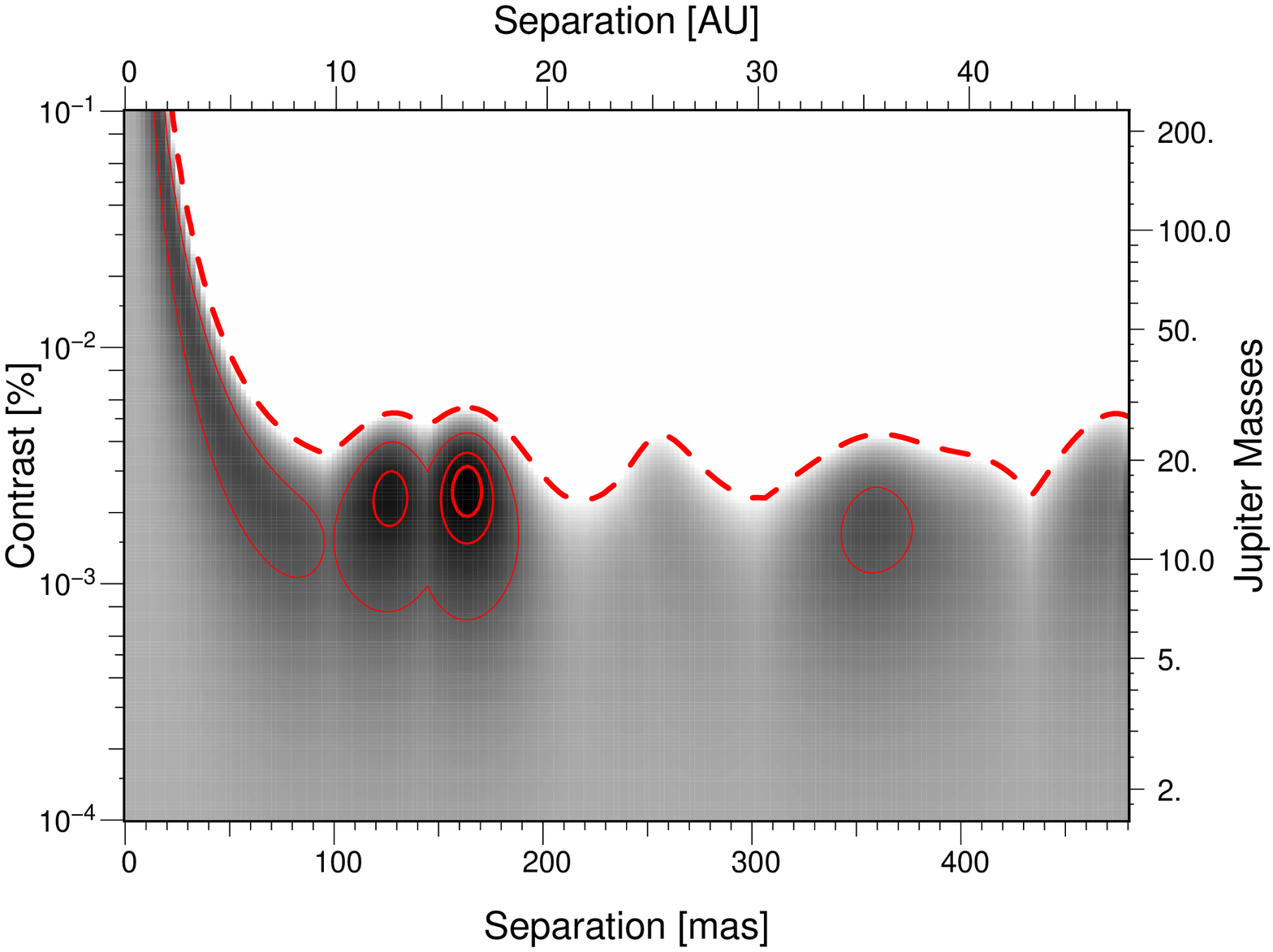}
   \caption{$\chi^2$ maps for target HD\,92945 (left panel) and
     HD\,141569 (right panel). The solid curves correspond to the
     isocontours at min($\chi^2$)+1; +4; and +9 (eg. one, two,  and
     three sigmas contours). Multiple minimums are present (false
     detections). The 5\,$\sigma$ detection limits are, averaged over
     the separation, $2.5\times10^{-3}$ and
     $4.6\times10^{-3}$ (dashed curve). The right vertical axis was
     obtained assuming a linear fit of DUSTY evolutionarily models.}
              \label{fig:nond}%
    \end{figure*}

\begin{table}
\caption{Best-fit detection}          
\label{chi2r}    
\centering     
\begin{tabular}{lccc} 
\hline\hline             
Target & Degrees of & $\chi^2$ &  Min $\chi^2$ \\
 & freedom &Single star&Binary model\\
\hline                      
 HD\,135549  &522&248318&908\\
\hline                      
 HD\ 92945   &137&282&234\\
\hline                      
 HD\,141569  &137&265&223\\
\hline                                   
\end{tabular}
\end{table}

Establishing a high-contrast detection limit follows a different approach 
than establishing a detection. The most important step is indeed to show
that there is no significant detection in our data. A practical
approach is to analyze the $\chi^2$ maps (Fig.~\ref{chi2_map_nd}).  If
the minimum value of the $\chi^2$ does not significantly differ from
the maximum $\chi^2$, it means that the binary model does not significantly
improve the fit to the data. The $\chi^2$ values at best fit 
are given in Table~\ref{chi2r}. The detection for the binary system 
HD\,135549 discussed in the previous section is extremely clear with 
a normalized $\chi^2$ of 1.85. This dramatically improves the
fit compared to the null hypothesis (no binary 
$\chi^2/n_{\rm freedom} = 423$). 
This is not the case for the two debris disk systems. At best fit, the
binary model improves the $\chi^2$ by a factor of 1.24 for HD\,92945
and 1.18 for HD\,141569.

Although the $\chi^2$ maps indicate a nondetection, robust
statistical limits based on these data remain to be derived. To
accomplish this, we plot the $\chi^2$ in  Fig.~\ref{fig:nond}, minimized
over position angle, as a function of separation
($\sqrt{\alpha^2+\delta^2}$) and contrast ($r$).  The red solid
contours correspond to 1, 2, and 3 $\sigma$ detection limits. Several
minima exist in the maps including two 3 $\sigma$ false detection: at
440\,mas (contrast 0.16\%) and at 160\,mas (contrast 0.2\%). It is
noteworthy to point out that the absence of a companion (equivalent to
a binary system with a separation of 0\,mas) is within 5 $\sigma$ of
the false detections. This 5\,$\sigma $ limit corresponds to the red
dashed curves. We define this curve as our 5\,$\sigma$ detection
limits, hence excluding any detection on the two dataset.  This
approach is coherent with the generally recognized 5$\sigma$ criterion
to validate a detection.

According to Fig.~\ref{fig:nond}, the average $5\,\sigma$ detection
limits of HD\ 92945 and HD\,141569 over the separation range
100--500\,mas are respectively $2.5\times10^{-3}$ and
$4.6\times10^{-3}$, that is, $\Delta L'$ of 6.5 and 5.8.  However, the
nondetection limit is not completely uniform over the field of view
and can have values that are nearly a factor of two worse.  We can observe the
detection limit worsening by a factor 2 between $\lambda/D$ and
$0.5\,\lambda/D$, with a precipitous drop for still lower spatial
separations. These results also confirm the Monte-Carlo simulations
(presented in Fig.~\ref{Monte}), assuming error bars on the closure
phase of 0.2 and 0.4 degrees.

\subsection{Jupiter-mass limits}

To convert contrast limits into upper bounds for the mass of any
possible companion, we must know the absolute magnitude corresponding
to the detection limit in each system.  We derived the L' band
magnitudes from the K band magnitudes of the calibrators: HD\,92933
(K=5.58\,mag) and BD-03\,3826 (K=6.70\,mag). We used the temperature
reddening from \citet{2000asqu.book..143T} to convert to L', and
scaled the magnitude with respect to the relative flux observed on
both calibrator and science star: HD\,92933 is $1.09\pm 0.03$ times
brighter than HD\,92945, and BD-03\,3826 is $1.11\pm0.05$ times
brighter than HD\,141569. As a result, we estimated the L' magnitude
of HD\,92945 to be 5.58\,mag and HD\,141569 to be 6.74\,mag. We
checked that these values are roughly compatible with the magnitude
derived solely by using the isochrones for pre-main sequence stars
\citep{2000A&A...358..593S} giving 5.6 and 6.3\,mag, respectively.
Scaling the brightness to derive the absolute magnitude at 10\,pc
(assuming distances stated in Sect.~\ref{targets}), we obtained
M$_{\rm L'}$=3.87\,mag (HD\,92945) and M$_{\rm L'}$=1.76\,mag
(HD\,141569).

Thus, the $5\,\sigma$ upper limits correspond to a nondetection of up
to a absolute magnitude of $M_{\rm L'}=10.4$\,mag for HD\,92945 and
$M_{\rm L'}=7.6$\,mag for HD\,141569.  To convert our detection limits
to planetary masses limits, we used {\it DUSTY} evolutionary models
\citep{2000ApJ...542..464C} convolved with the NaCo filters. Towards
HD\,92945, the models put a limit on the mass of a companion to
18~\Mjup\ at a separation of 1.5\,\au ($\lambda/D$). Towards HD\,141569,
the models limit the mass of a companion to 22~\Mjup\ at
7\,\au.

\section{Conclusion}

 We have shown that aperture masking gives detection limits of the
 order of $\Delta~{\rm L'\,mag}=6$, with an inner working angle close
 to $\lambda$/2D. These results confirm previous detection limits
 obtained by the same aperture masking technique on the Keck telescope
 \citep{2011ApJ...731....8K,2011ApJ...730L..21H} In terms of
 scientific impact, this observational domain is important because it
 corresponds to a few astronomical units at a hundred parsec, the
 distance where the closest formation regions are. The scientific
 importance of this parameter space is highlighted by T Cha b detected
 by the same technique \citep{2011A&A...528L...7H}.

  Simulations of the older debris disk tend to show that a
  magnitude or two in dynamic range is still needed to observe disk
  shaping planets. Considering HD\,141569, for example,
  \citet{2009A&A...493..661R} show that the disk geometry could be
  best modeled by a flyby star and a planet of a few Jupiter masses. For
  these kinds of objects, a detection limit of two Jupiter masses would
  require a precision on the closure phases of 0.01 degree, something
 only  possible if we understand how to precisely account for the
  systematics errors.

This paper has to be considere along with other direct detection
techniques.  In the case of HD\,92945, \citet{2007ApJS..173..143B}
used a spectral differential technique to derive an upper limit of
10.8 mag at 0.5'' in the H band (5\,$\sigma$). This limit assumes a
clear methane signature (a spectral type T8 for the planet) which may
not be the case. This assumption is not required for a differential
technique using sky rotation: in their Gemini survey,
\citet{2007ApJ...670.1367L} obtain the same dynamic range (10.8 mag;
5\,$\sigma$) at the slightly larger separation of 0.75''. Overall,
differential imaging techniques often give a dynamic range that is
higher than SAM, but with larger inner working angles (typically a few
$\lambda$/D).

The capacities of future extreme adaptive optics to observe at smaller
inner working angle, compared to the ability for aperture masking to
probe at higher dynamic range by understanding the systematics, will
decide the fate of SAM with respect to faint companion detection.

\begin{acknowledgements}
SL would like to acknowledge the friendly support of ESO's staff
during observations, including night time support astronomers and
telescope operators. A special thanks goes to Dr. J. Girard, the
instrument scientist of NaCo.  This work received the support of
PHASE, the high angular resolution partnership between ONERA, the
Observatoire de Paris, CNRS, and University Denis Diderot Paris 7.
\end{acknowledgements}

\bibliographystyle{aa}   
\bibliography{sambib}   

\end{document}